\documentclass[draft]{article}
\usepackage{latexsym}
\usepackage{graphics}
\usepackage{epsfig}
\usepackage{rotate}
\usepackage{amsmath}
\usepackage{amssymb}
\usepackage{array}
\usepackage[dvips]{color}
\usepackage{afterpage}
\usepackage{graphics}
\newcommand{\beq}{\begin{eqnarray}}
\newcommand{\eeq}{\end{eqnarray}}
\newcommand{\sop}{standard oscillation phase }
\newcommand{\sol}{standard oscillation length }

\begin{document}
\title{On ultra-relativistic approximations, unobservable phases and other hand-waving in the derivation of the neutrino 
oscillation length.}
\author{J.-M. Levy \thanks{Laboratoire de Physique Nucl\'eaire et de Hautes Energies,
CNRS - IN2P3 - Universit\'es Paris VI et Paris VII, Paris.   \it Email: jmlevy@in2p3.fr}}
\maketitle
\begin{abstract}

A wrong derivation of the phase of a propagating massive particle which has repeatedly appeared during the last years has
the advantage of leading once more to the important question of the phase difference between the mass eigenstates
constitutive of an oscillating neutrino described in the plane wave formalism. Serious errors of principle are pointed at
in a number of simple calculations and the oscillation length is derived in a way which allows to show that the standard
result can suffer but minute variations in the ultrarelativistic case.

\end{abstract} 
\newpage 
\section{Introduction}

Neutrino oscillations now seem to be firmly established on experimental grounds. It is therefore of necessity to know
how to extract physical parameters, viz masses and mixing matrix elements, from the data. Of paramount importance for
this purpose is the formula linking the measured oscillation parameters to the unknown masses of the eigenstates
constitutive of an oscillating neutrino. This famous 'oscillation length formula' has been challenged by many authors
and its derivation is often bungled by others. The purpose of this paper is not to challenge the 'orthodox' point of
vue, but to try to derive the oscillation phase as neatly as possible in the plane-wave formalism, baring hand-waving
and faulty approximations.  It is not a difficult exercise, but it is, curiously enough, at variance with a number of
papers by authors who, knowing the result in advance, manage to find the right answer for the wrong reasons, doing 
errors which are often the same as those of the proponents of alternative formulae and are based on the same ill 
employed approximations. Prominent among these are the so-called ultra-relativistic approximations which amount, for the
careless, to the mere replacement of $v$ by $c=1$ without any consideration for what they were taught about doing
asymptotic expansions. The first purpose of this paper is therefore pedagogical. It is deeply deplorable that so many
wrong derivations are still plaguing seminars, articles and books on this question, especially when papers appear in
journals devoted to the teaching of physics.\\ However, this article is also aimed at physics itself. We hold it that 
doing proper maths helps to understand the problem. That the phase is not what the authors of wrong papers want it to be
forces to employ a little more thought; there is always something to be learned from such exercices. With very 
elementary maths, we end up showing that there is no room for deviations from the \sol with light, ultrarelativistic
neutrino-like particles, that is, in the only case where an oscillation length has a physical meaning.

\section{Phase of a definite mass particle}\label{int} 

The derivation of the phase of a propagating massive neutrino published in \cite{Giunti:2003ku} (see also
\cite{Giunti:2003tp,Giunti:2003qt,Giunti:2004pd,Giunti:2004yg,Giunti:2004zf,Giunti:2006fr,Giunti:2008cf,PDBook,Yao:2006px}
and the book \cite{Giunti:2007ry} p.290 eqn 8.37 and probably many more which reproduce the same calculation)  is plainly
wrong.\\

Starting from the relative phase \footnote{A $c = \hbar = 1$ system of units is used} \beq \Phi(L,T) = pL-ET
\label{phase}\eeq acquired by the state vector of a particle of energy $E$ and momentum $p$ upon displacement from
space-time point $(0,0)$ to space-time point $(T,L)$, the author makes the 'ultra-relativistic approximation' $T = L$ on
the ground that neutrinos are ultra-relativistic particles, and from there, using the dispersion relation $E^2=p^2+m^2$
and $p \approx E$ for an ultrarelativistic particle, he gets the following equation:
 $$\Phi = (p-E)L = -\frac{m^2}{E+p}L\approx  -\frac{m^2}{2E}L \;\;\;\;\;\; (a)$$ 

We claim that $pL-ET = (p-E)L$ is an ugly equality which should be banned from decent physics litterature although it
can be found in many places, starting with the review dedicated to neutrino masses and oscillations in various editions
of the celebrated Review of Particle Physics (see eg \cite{PDBook},\cite{Yao:2006px} )\\

It is quite clear that this cannot be correct:\\ 
\begin{itemize}

\item First and foremost, $L$ is $vT$, not $T$ and replacing the variable by its limit in calculating an asymptotic
expansion is not allowed. If he were self-consistent, the author of \cite{Giunti:2003ku} would do the same with $p = vE$,
afterwards the phase would collapse..

\item Second, the result must fit seamlessly with what is obtained without the ultra-relativistic approximation. But then,
when $v < 1 $, anybody would use this obvious $L= vT$ which, when combined with the $p = vE$ relation also valid for any
velocity, yields \beq\Phi = ET(v^2-1) = \frac{-m^2T}{E} = \frac{-m^2L}{p} \label{true}\eeq This is equal to twice the
expression found in \cite{Giunti:2003ku} (see eqn. (a)) in the $v=1$ limit, and there is nothing to tell us when, and for
which bizarre reason, one should suddenly drop this factor when $v$ gets close to one.  

\item Third, Lorentz invariance dictates precisely what the phase of a plane wave must be in an arbitrary frame.  Any wave
equation yields the Schr\"odinger time dependence $exp(-i m \tau)$ in the rest frame \footnote{$\tau$ is the proper time}
and consequently the phase in any frame is given by (\ref{true}), not by half of it. Here $L$ is understood as the spatial
distance traveled by a classical particle of energy $E$ and linear momentum $p$ in a time $T$. We shall give below a proof
of this Lorentz invariance which allows to demonstrate (without approximations) the point that the author of
\cite{Giunti:2003ku} wanted to make in his paper; but it must be stressed that Lorentz invariance of the phase is a
logical requirement as basic as invariance of the numbers of nuclei in a given lump of matter.  Although the author of
\cite{Giunti:2003ku} claims that (\ref{phase}) is Lorentz invariant \footnote{Of course, Lorentz invariance of
(\ref{phase}) has no meaning as long as $L$ and $T$ or $E$ and $p$ are not somehow related.}, he does not seem to mind the
fact that his 'approximation' yields only half the Schr\"odinger phase in the rest frame.

\end{itemize}

Curiously, a similar treatment employed in the Review of Particle Physics has been challenged, not because of its
physical and mathematical unsoundness, but on the ground that the 'overall' phase is not observable \cite{Lipkin:2005kg}
and is not gauge-invariant and that readers were induced by the text into taking "the phase seriously". It is clear
however, that this is not an 'overall' phase, but the phase difference acquired in a specific space-time translation and
that it is precisely this phase which is used when building a wave packet or deriving a space-time propagator from the
path integral. Since it is a difference, it is obviously gauge-invariant (whatever the relevance of this concept for a
free non-interacting neutrino..) and should be "taken seriously" as it is done in, e.g., any derivation of the solutions
of a wave equation starting from the rest frame. Anyway, even non gauge-invariant constructs such as the electromagnetic
potentials have to be taken seriously as far as basic maths are concerned. Their inobservability does not justify
hand-waving calculations. In the case at hand, we all know that there is no such thing as an intrinsic (rather than 
"overall") phase of the state vector, but phase variations do have a meaning and cannot be multiplied by arbitrary 
factors.\\

Observe also that no justification is givent in \cite{Giunti:2003ku} for the use of an 'ultrarelativistic approximation'
($v$=1) in the relation between elapsed time and distance and its not being used in the relation between energy and
linear momentum.  The reason for using the first approximation is evidently that it later allows to calculate the phase
difference (between two mass eigenstates) at a unique same space-time point without further discussion; the unstated
reason for not using the second approximation is that the phase would collapse to $0$ as for light rays, if both were
used simultaneously.  (This second approximation is in fact also used, but only in the very last step of the derivation,
when $p$ is safely added to $E$ instead of being subtracted..)\\


Therefore, the reason for the difference and the factor of two boils down to $1-v^2 \approx 2(1-v)$ when $v
\rightarrow 1$. Using only one factor of $v$ amounts to dropping the 2.\\

The same trick is used with the same lack of care and comments in many such 'simple' derivations of the oscillation
length. Physicists should know that an equivalence is not an equality, that a statement about a limit is a particular
kind of equivalence and that even a physical quantity never becomes equal to its limit under any approximation when
this limit is outside the physical range. 'Equal in the ultrarelativistic regime' is devoid of meaning and fraught
with traps.


\section{Lorentz invariance of the phase} \label{lori} 

Note, however, that we do not challenge the main point of \cite{Giunti:2003ku} viz. the Lorentz invariance of the
oscillation probability, which must hold for obvious consistency reasons. We shall start by showing, along the lines of 
\cite{Giunti:2003ku} but with a little more care and without any 'ultra-relativistic approximation', that
the Lorentz invariant quantity is really $T/E = L/p$ where $(L,T)$ are the space-time coordinate differences between
the events of production and detection. Contrary to the claim of \cite{Giunti:2003ku}, his 'approximation' $T=L$ cannot be 
the crucial point in a check of Lorentz invariance, which must also hold between two frames wherein the particle 
velocities are not close to $c$ \\

Let us assume that the object is produced at point $x=0$ and time $t=0$ and that its speed is $V$ in frame {\it K} where
the stationnary target is situated at $x=L$. Let $T=L/V$ be the impact time in {\it K}.\\ In a frame $\it K'$ with axes
parallel to those of $\it K$ and moving with velocity $v$ along $0x$, with the origins so choosen that $t'=t=0$, the
impact time will be $T' = \frac{T-vL}{\sqrt{1-v^2}} = T \frac{1-vV}{\sqrt{1-v^2}}$ On the other hand, the object energy in
$\it K'$ will be $E' = \frac{E-vp}{\sqrt{1-v^2}} = E\frac{1-vV}{\sqrt{1-v^2}}$ if its energy and momentum are $E$ and $p$
in $\it K$. Therefore $T'/E' = T/E = L/p = L'/p'$ \footnote{The elementary derivation just given can be mimicked for $L$
and $p$. The second equality trivially follows from $p=vE$ and $L=vT$} as announced and this immediately implies Lorentz
invariance of the phase as shown by formula (\ref{true}).  The true invariant is $ T/E = L/p$~ and not, as stated in
\cite{Giunti:2003ku}, $L/E$ to which it is {\bf equivalent} (not equal !) in the ultrarelativistic regime.  For further
reference, observe that the equality holds because the same $V$ is used in both position and momentum space.

\section{What is the "equation of the trajectory" ?}
We are dealing with quantum objects and one might criticize our use of $v=p/E$ in the time-distance relation; 
however, it is well known that, when one uses wave packets instead of plane waves, the center of the free 
wave packet follows the classical trajectory, this being true both in non-relativistic and relativistic quantum mechanics 
\footnote{In the latter case, and depending on how the wave-packet is built, there is a quite negligible  additional 
oscillation around the central value, (zitterbewegung), with an 
amplitude of the order of $\hbar/E$ that is, $2 fm$ for a 10 MeV particle}. The corresponding velocity
is the  group velocity of the packet as can be shown by a stationnary phase argument or by 
calculating the expectation value of the position operator. Using $v = \frac{dE}{dp} = \frac{p}{E}$ 
in the plane wave approximation where the time-distance relation must be added by hand is therefore well 
justified. Hence, if a phase is to be used for the center of the wave packet describing a one particle state, it 
cannot be anything else than (\ref{true}). But we shall see that it is quite irrelevant for the oscillation problem.

\section{Phase difference of different mass eigenstates} 
\subsection{Standard approximations} 

Standard derivations of the oscillation formula in the plane-wave formalism resort most often to approximations known 
as "equal $E$" or "equal $p$" in which one postulates equal energies or equal momenta for the two mass-eigenstates 
constitutive of the flavour neutrino, notwistanding the contradiction with production kinematics and the fact that 
these hypotheses are quite arbitrary - these equalities aren't even Lorentz invariant. The reason for so doing is that 
a connection between time and space must be postulated since it cannot be inferred in this case from a dynamical 
calculation. With the above mentionned approximations, the spatial or the temporal interval disappears upon 
subtracting the phases of the two mass eigenstates and there is no discussion about the (space-time) point where one 
chooses to calculate the interference pattern. Specification of the distance is enough. Using his above equation $(a)$ 
in the so-called "equal $E$ approximation", the author of \cite{Giunti:2003ku} immediately falls back on the standard 
result that gives rise to the neutrino oscillation length quoted in hundreds of papers on the subject. There are, 
however, less questionnable ways of retrieving the same result. \subsection{What is wrong with these phases ?} As we 
have shown, the phase is given by (\ref{true}) and at first sight, subtracting similar expressions written for $m=m_1$ 
and $m=m_2$ will yield twice the standard result for the oscillation phase.\\ To get their factor straight, the 
authors of \cite{Okun:2002ye} introduce a "space velocity $v_s$" different from the phase velocity and write $\Phi = 
-ET(1/\gamma^2+1/\gamma_s^2)/2$ after what $\gamma_s$, which is the same for all mass eigenstates, cancels in the 
substraction, yielding the requested factor 1/2. Indeed, $v_s = L/T$ is the same for all mass eigenstates "as required 
by quantum mechanics" write the authors of \cite{Okun:2002ye}. \\

We prefer a less contorted approach, using directly the already established results. A same "space velocity" would
preclude the separation of the centers of the wave packets and the washing out of oscillations which will necessarily
occur in a complete treatment of the problem.\\ However, sticking to what we have so far established, the result
(\ref{true}) does not mean that the \sop is wrong and that proponents of the formula modified by the famous factor of 2
are right.  The point is that for different values of $m$ (and a given observation point, that is, a given $L$ in a
frame were the target is stationnary) the factor $T$ in (\ref{true}) cannot be the same. Use has been made of the
kinematics ($T = L/v$)  of the center of the wave packet and since the energies are supposedly equal, different masses
lead to different velocities, hence to different travel times. Likewise, using the second form of $(2)$, the equal $L$
factors hide different times because of different velocities. But phases must be compared at the same space-time point.
It is therefore meaningless to subtract them after having reduced them to the form (\ref{true}).\\

It is quite obvious that there is no way to derive any interference from classical kinematical calculations.  
Classical particles are concentrated, so to say, at the center of their wave packets and therefore cannot interfere as
soon as these centers separate. On the other hand, what is required by quantum mechanics is to calculate overlap
amplitudes at a given point and a given time.  Using different instants and/or different space points simply makes no
sense. This is the true justification for the same "space velocity" of \cite{Okun:2002ye} and this is where most
proponents of alternative formulae stumble. They forget that elementary particles have little to do with sand grains and
that it is only the centers of the wave packets which follow $L=vT$ trajectories.  But it is precisely that "fuziness"
which allows to calculate a phase difference and to find interferences and oscillations.  Classical particles cannot
interfere. 
\subsection{Various solutions}
The solution here is to go back to the original expression (\ref{phase}) of the phase where $L$ and $T$ can be fixed
independently.  Then, in the equal $E$ approximation, the time-energy parts cancel and the space-momentum parts yield
the answer: $$\delta \Phi = Lp_1-Lp_2 = \frac{L(p_1^2-p_2^2)}{p_1+p_2} = -L\frac{m_1^2-m_2^2}{2\bar{p}}$$ where use has
been made of the 'equal $E$' hypothesis and an obviously defined $\bar{p}$ has been introduced. Note that we have not
used any relativistic approximation to expand the momenta. The expression found is valid in principle for hypothetical 
non relativistic neutrinos (\cite{Levy:2000nv})\footnote{See however the next section for a caveat}.

The same reasoning applies to the "same p" approximation. Formula (\ref{true}) written in terms of $p$ and $L$ does not
lend itself to a simple subtraction between mass eigenstates, because the two subtracted expressions do not correspond
to the same time for the already explained reasons. Here again, one must go back to the original expression of the
phase, whereoff the $pL$ parts cancel upon subtraction to yield: $$\delta \Phi = TE_2-TE_1 = 
T\frac{E_2^2-E_1^2}{E_1+E_2}=T\frac{m^2_2-m^2_1}{2\bar{E}}$$ The last expression must be converted to a function of 
distance to make connection with experiments. Using the average velocity defined by $p/\bar{E}$, one retrieves
the expression found above with $\bar{p} \rightarrow p$, again without a relativistic expansion. (see 
\cite{Levy:2000nv}) 

Even the (notoriously unphysical, see \cite{Okun:2000gc}) "same velocity" approximation yields the standard answer 
when treated with care \cite{Levy:2000nv}. The reason for which a wrong hypothesis leads to a reasonable result
will become clearer in the next section.

\section{A better Ansatz} \label{ansatz}

In \cite{Levy:2000nv, Levy:2000qx} the present author (unaware of the existence of \cite{Okun:1999se}) showed that not
only could one retrieve the standard result without any relativistic expansion as shown here above, but also without any
of the "equal energy" or "equal momentum" or "equal velocity" approximations, by postulating an appropriate velocity for
the center of would-be wave packet.  This velocity is defined as $v_0 = \frac{p_1+p_2}{E_1+E_2}$ (see also
\cite{Burkhardt:1998zj} and \cite{Burkhardt:2003cz} where this same value is used to disprove the existence of
oscillations for the recoil particle) It was guessed on the basis of the average velocity found as already described in
the equal $p$ approximation, or in the equal $v$ approximation were it is derived from the (by itself unrealistic)
assumption that $\frac{p_1}{E_1}=\frac{p_2}{E_2}$ \footnote{The reason for which this hypothesis is wrong in principle is
that strictly equal velocities entail masses in the same ratio as energies that is, almost equal for a beam of more or
less defined energy - and equal masses are obviously completely unrealistic (see \cite{Okun:2000gc}). The practical reason
for which it works is that the inverse function $\gamma -> v =\sqrt{1-1/\gamma^2}$ is flat and undistinguishably close to
1 well before $\gamma$ reaches values which are relevant to the neutrino case. Therefore very different $E/m$'s yield the
'same' (for all practical purposes) velocities} \\
 
As already said, the usual equality assumptions contradict production kinematics, be it through a reaction or through a
decay.  No such violence is done to kinematics in the present case. Energy and momenta can be given their correct,
conservation laws respecting values. The particular expression for $v_0$ arises quite naturally from writing the phase
difference as: $$\delta \Phi = x \delta p - t \delta E = x \frac{\delta p^2}{\Sigma p} - t \frac{\delta E^2}{\Sigma E}$$
Converting this to a function of $x$ alone, one is led to write $t = \frac{x}{v}$ and the choice $v = \frac{\Sigma
p}{\Sigma E}$ tends to impose itself, yielding: $$\delta \Phi = -x\frac{\delta m^2}{\Sigma p} = -x\frac{\delta
m^2}{2\bar{p}}$$ which is the \sop with the bonus of showing that the kinematical variable to use, would the distinction
between energy and momentum be relevant, should be the latter, since this calculation is valid for any velocities 
(\cite{Ahluwalia:1997tr}, \cite{Levy:2000nv, Levy:2000qx})  
\footnote{Note however, that an oscillation probability independent of the production and detection processes is
an illusion for hypothetical neutrino-like particles of sizeable mass. No factorization of production, propagation and 
detection probabilities can hold here}.\\

Although natural in the above context, the value $v_0$ springs out of the blue and is not justified by any convincing
calculation.  It should come out of the dynamical evolution equation in a wave packet or field-theoretical treatment,
but since our purpose is to keep things simple, we will not embark on such calculations and refer the reader to the
enormous existing litterature which we don't even feel able to review properly. For a very complete report of elaborate
treatments, see \cite{Beuthe:2001rc}. Here, we will content ourselves with showing that deviations from this 'desirable'
velocity, as Lev Okun calls it, do not lead to any serious alterations of the \sol.

\section{Oscillation length}

$v_0$ as above defined can trivialy be looked at as an average of the classical velocities of the two mass eigenstates
weighted by their energies: $$v_0 = \frac{E_1}{E_1+E_2}v_1 + \frac{E_2}{E_1+E_2}v_2$$ with $v_i = \frac{p_i}{E_i}$. Hence
it lies between these two, like any reasonable value for the classical velocity of a mixed state. We shall therefore
parametrize an arbitrary (but 'reasonable') velocity by $1/v = 1/v_0 + \epsilon$ which yields the phase shift as a
function of distance : $$\delta \Phi = -x\frac{\delta m^2}{2\bar{p}} - x\frac{\delta E^2}{\Sigma E} \epsilon =
-x\frac{\delta m^2}{2\bar{p}} - x\frac{\delta E^2}{2\bar{p}}v_0\epsilon $$ yielding an oscillation length: $$ L =
\frac{4\bar{p}\pi}{\delta m^2 + v_0 \delta E^2 \epsilon} = \frac{4\bar{p}\pi}{\delta m^2 + 2 \bar{p} \delta E \epsilon}$$
which differs from the standard result by the second term in the denominator.\\

Assume now that the neutrino is produced in a reaction with total c.o.m. energy squared $s$. Then $E_i =
(s+m_i^2-m_r^2)/(2\sqrt{s})$ where $m_r$ is the effective mass of the system recoiling against the neutrino. 
Then $\delta E = \delta m^2/(2\sqrt{s})$ and we have: $$ L = \frac{4\bar{p}\pi}{\delta m^2(1+\bar{p}\epsilon/\sqrt{s})} 
= \frac{L_0}{1+\bar{p}\epsilon/\sqrt{s}}$$ where $L_0$ is the \sol.\\

Now for light neutrinos, $\bar{p} \approx \bar{E} \approx 1/2(\sqrt{s}-m_r^2/\sqrt{s})$; $\bar{p}/\sqrt{s}$ is less than 
1/2 and can be fairly lower. For example, in a $\pi \rightarrow \mu \nu$ decay, it is about $.21$ and it can be still 
much lower in three body decays.\\

$|\epsilon|$ itself can be easily bounded from its definition. But it clearly can't exceed $|v_1-v_2|/(v_1v_2)$ 
which is extremely low since with standard neutrino masses, many figures will cancel between the two $v$'s in the numerator.
 Therefore there isn't much room. Only very small numerical variations w.r.t. the \sol value are possible in this frame
\footnote{At $10\;GeV$ c.o.m. energy, a $\nu$ assumed to have an $19\;MeV$ mass (limit derived for $\nu_{\tau}$'s from
$\tau$ decays but enormously larger than expected standard neutrino masses given the measured $\delta m^2$'s and the 
limits
on $m_{\nu_e}$) recoiling against a proton has $v = .99999x$ where $x$ stands for the first digit not equal to $9$}. One
sees that the reason which makes this oscillation length so sharply defined is the same that renders the equal velocity
hypothesis practically useable in spite of its being wrong in principle. Indeed, $v=1$ after all, even if it is not
permitted to write it before the end of the calculation !

\section{Lorentz invariance of the oscillation phase}

Be it calculated with the usual equality hypotheses or with the ansatz proposed in section \ref{ansatz} for the classical
velocity of the mixed neutrino, the oscillation phase is:  $\delta \Phi = -x\frac{\delta m^2}{2\bar{p}}$ and it is Lorentz
invariant, as shown in section \ref{lori}, as soon as $x$ is interpreted as the distance traveled by the neutrino between
production and detection in the frame where it has momentum $\bar{p}$. It has to be so since this phase directly gives the
probability that an interacting neutrino yields a lepton of the initial flavour.  The 'oscillation length' $L_{osc}$,
derived by equating this phase to $2\pi$ has the meaning of the source-detector distance giving a maximum of this
probability only in the frame where the target is stationnary. In other frames it can be found by scaling $L_{osc}$ in the
same ratio as the momenta and it represents the distance traveled by the neutrino during its existence.


\section{Conclusion} Having recently assisted to a seminar where the untenable derivation of the oscillation phase as
above described was proposed, the present author has decided to write this note in the hope of contributing to bar
widespread and very unpedagogical hand-waving in the simple way of attacking the question of vacuum neutrino oscillations.
He took the occasion to show why only minute deviations from the \sop are possible in this simplified frame and to 
straighten up the derivation of the Lorentz-invariance of the oscillation probability. \\

\bibliographystyle{epj}
\bibliography{Phase_of_neutrinos}
\end{document}